\documentclass[twocolumn,showpacs,preprintnumbers,amsmath,amssymb]{revtex4}

\usepackage{graphicx}
\usepackage{dcolumn}
\usepackage{bm}

\newcommand{\ket}[1]{\left| #1 \right>}
\newcommand{\bra}[1]{\left< #1 \right|}

\begin{document}
\title{ Reversible entanglement in a Kerr-like interaction Hamiltonian: an 
integrable model}
\author{L. Sanz, R.M. Angelo and K. Furuya}
\affiliation{Instituto de F\'{\i}sica `Gleb Wataghin',
Universidade Estadual de Campinas, Unicamp\\ 13083-970, Campinas,
SP, Brasil}
\date{\today}

\begin{abstract}
An exactly soluble non-linear interaction Hamiltonian is proposed to study
fundamental properties of the entanglement dynamics for a coupled
non-linear oscillators. The time-evolved state is obtained
analytically for initial products of two coherent and two number states and 
relevant informations are extracted from the dynamics of various quantities 
like subsystem linear and Von Neumann entropies, quadrature mean values,
variances and $Q$-functions. We determined the re-coherence
time scales and found among the interaction terms
present in the Hamiltonian the one responsible for the entanglement
in both cases. We identify the existence of two regimens
for the entanglement dynamics in the case of initially
coherent states: the short time, {\em phase spread} regimen where the 
entropy rises monotonically and the {\em self-interference} regimen 
where the entropy oscillates and re-coherence phenomenon can be observed. 
We also found that the {\em  break time} from the first regimen to the second 
one  becomes longer, as well as the re-coherence and reversibility times,
as the Planck's constant becomes much smaller than 
a typical action in phase space. \\

\end{abstract}
 \maketitle
\newpage
\vspace{\bigskipamount}
\section{\label{sec:intro}Introduction}

Nowadays it is widely accepted that quantum entanglement is an
essential ingredient for the implementation of quantum information
processing devices \cite{werner}. It is notable that until recently
quantum information science was restricted over discrete, finite
dimensional Hilbert space elements, however, interest on
continuous variables has been developing in various contexts:
from teleportation \cite{braunstein98}, cryptography \cite{ralph} to cloning
\cite{braunstein01}. This is closely connected to the recent advances of
enhancing nonlinear coupling via electromagnetically induced
transparency (EIT) mechanism \cite{schmidt} using a Bose-Einstein condensate, 
which has opened possibilities of strong non-linear interaction of ultra 
slow light pulses \cite{hau} of tiny energies belonging to different modes 
of electromagnetic fields \cite{lurkin,deng}. 
Such advances ranges from building a quantum logic gates based on 
photonic qubits without resorting to photonic crystals \cite{petrosyan}, 
entangling continuous variable states \cite{braunstein98,silberhorn}, and 
preparing entangled states of radiation from mixed thermal states \cite{filip}
in conventional media. Facing these recent developments, it is
important to understand the entanglement properties of Kerr-type
interaction particularly for Gaussian states, which are one of the states
that can be generated by simple experimental devices like beam splitters
and phase shifters.

In the present work, we have studied the entanglement properties of a 
model describing a bipartite system of two degrees of freedom, an integrable 
version of two quartic oscillators \cite{milburn86,agarwal98}, coupled via a 
Kerr-type interaction. One could think of some coupled cavities scheme
with a Kerr medium \cite{scully,opatrny} as a possible realization of these 
type of interactions. The nonlinearity of each of the subsystem oscillator 
keep some of the  properties present in one degree of freedom such as 
collapses and revivals of the quadrature mean values of each field. 
Also, this soluble model in which the nonlinearity is present can be used to
understand the role of Kerr-type  nonlinearity on the entanglement
dynamics of initially coherent Gaussian states. Another case we
have considered is the non-classical number states \cite{brune} for the initial
disentangled state, which does not entangle via Kerr-type coupling, but via 
usual bilinear coupling in a rotating wave approximation (RWA), 
and show several differences in the entanglement dynamics.

 The paper is organized as follows: In section II the model is introduced 
and the exact solutions for initially disentangled states, in the cases of
number states and coherent states, are presented. Analytical expressions 
for the subsystem density operators and some mean values are calculated. 
Section III is reserved to present the exact subsystem linear entropies for 
both initial states, analyze the conditions for re-coherences and 
recurrences and discuss the differences in the entanglement dynamics in 
the proposed cases. 
Also, we examine the differences of the short time dynamics where no 
interference phenomenon is possible and the longer time dynamics where such 
effects do affect the evolution of the subsystems. Another issue treated 
here is the limit where Planck's constant becomes much smaller than the 
typical action in phase space. As we shall see, the two cases present 
distinct semi-classical behaviors. Finally in section IV we present our 
conclusions.

\section{\label{sec:model}The model}

Our theoretical model will be inspired on two field modes in a cavity
with a low-loss Kerr media, described by the creation and annihilation
operators $\hat{a}_k$ and $\hat{a}^{\dagger}_k$ ($k=1,2$), such that
each mode has the usual non-linear interaction term of the form $\chi
\hat{a}^{\dagger 2}_k {\hat{a}_k}^2$ \cite{agarwal98}. Besides this non-linear 
self-interaction term we add two coupling interaction between the field
modes: one is the usual RWA coupling \cite{mokarzel} and the other is the 
Kerr-type non-linear interaction \cite{scully},
\begin{eqnarray}
\hat{H}_{eff}&=&\sum\limits^{2}_{k=1}\left[\hbar\omega_k 
\left(\hat{a}^{\dagger}_k\hat{a}_k+\frac{1}{2}\right)\right]+
\hbar\lambda \left(\hat{a}_1^{\dag}\hat{a}_2+\hat{a}_1\hat{a}_2^{\dag}\right)
\nonumber\\
&&+\hbar^2 g \left(\hat{a}_1^{\dagger 2}{\hat{a}_1}^2+
                   \hat{a}_2^{\dagger 2}{\hat{a}_2}^2 \right)+
 \hbar^2 g^{'}\hat{a}_1^{\dagger}\hat{a}_1\hat{a}^{\dagger}_2\hat{a}_2.
\nonumber
\end{eqnarray}
where the constants $g$ and $g^{'}$ \cite{18} depend upon the third-order 
non-linear susceptibility $\chi$  and no losses will be considered. 
Since our purpose here is to have an exactly soluble model with the
nonlinearities in such a way that we could trace out the effect of
each interaction term on the entanglement dynamics, we will just
set $g^{'} = 2 g$ and also choose the resonant case ($\omega_1=\omega_2$) 
such that the Hamiltonian can be re-written in the following form:
\begin{eqnarray}
\hat{H}&=& \hbar\omega_0 \sum\limits^{2}_{k=1}\left[
\left(\hat{a}^{\dagger}_k\hat{a}_k+\frac{1}{2}\right)\right]+
\hbar\lambda \left(\hat{a}_1^{\dag}\hat{a}_2+\hat{a}_1\hat{a}_2^{\dag}\right)
\nonumber\\
&&+\hbar^2 g \left(\hat{a}_1^{\dagger}{\hat{a}_1}+
                   \hat{a}_2^{\dagger}{\hat{a}_2} +1 \right)^2\nonumber\\
&=& \hat{H}_0+\hat{V_{\lambda}}+\hat{V_{g}}.
\label{eq:hamiltonian}
\end{eqnarray}
Notice that the entire Kerr type interaction is now in the form of the 
square of the free Hamiltonian
$\hat{H_0}$, and hence the following commutation relations hold\\
\begin{eqnarray}
\left[\hat{H}_0,\hat{V_{\lambda}}\right]=\left[\hat{V}_g,\hat{V_{\lambda}}
\right]=0. \label{eq:commutator}
\end{eqnarray}
This property is important for two reasons: first, it is associated to
the presence of a constant of motion ($\hat{\mathcal{N}}=
\hat{a}^{\dag}_1 \hat{a}_1+\hat{a}^{\dag}_2 \hat{a}_2$); and second, it 
allows us to study separately the action of each interaction
term in the evolution operator on the initial state, and hence its
consequences on the entanglement process. 
In what follows, we study the cases of initially uncorrelated  number and 
coherent states. 
\subsection{\label{sec:solnumber} Analytical solution for product of number 
states}
\noindent
For the case of initially disentangled number states of the harmonic 
oscillators:
\begin{equation}
\ket{\psi(0)}=\ket{n_1}\otimes\ket{n_2}. \label{eq:ininumber}
\end{equation}
We can write $\hat{H}$ in Eq.(\ref{eq:hamiltonian}) in terms of
two new quartic oscillators diagonalizing the RWA-coupling, using the
transformation proposed by Zoubi {\it et al}~\cite{zoubi},
\begin{eqnarray}
\begin{array}{c}
\hat{a}_1=\frac{\hat{A}_1+\hat{A}_2}{\sqrt{2}} \qquad \qquad
\hat{a}_2=\frac{\hat{A}_1-\hat{A}_2}{\sqrt{2}}. \label{eq:trans}
\end{array}
\end{eqnarray}
The resulting Hamiltonian can be written in terms of the number
operators $\hat{N}_k=\hat{A}_{k}^{\dag} \hat{A}_k$, of the {\em
new} oscillators as follows:
\begin{eqnarray}
\hat{H}&=&\hbar \left( \omega_0+\lambda\right) \left(\hat{N}_1+
\frac{1}{2}\right)+\hbar \left( \omega_0- \lambda\right)\left(\hat{N}_2+
\frac{1}{2}\right)\nonumber\\
&&+\hbar^2 g \left( \hat{N}_1+\hat{N}_2+1 \right)^2.
\label{eq:Hnum}
\end{eqnarray}
Using Eq.(\ref{eq:trans}), we can connect the two different basis of the
two oscillators Hilbert space by the relation

\begin{equation}
{\small
\ket{n_1,n_2}_a=\sum^{n_1}_{i=0}\sum^{n_2}_{j=0}
c_{i,j}(n_1,n_2)
\ket{\mathcal{N}-(i+j),(i+j)}_A},
\label{eq:basis}
\end{equation}
where
\begin{eqnarray}
c_{i,j}(n_1,n_2)&=&(-1)^j{n_1\choose i}{n_2\choose j}
\sqrt{\frac{[\mathcal{N}-(i+j)]!(i+j)!}{2^{\mathcal{N}}n_1!n_2!}},\nonumber\\
\mathcal{N}&=&
\left<n_1,n_2|\hat{\mathcal{N}}|n_1,n_2\right>=n_1+n_2.
\end{eqnarray}
The sub-index `$a$' or `$A$' indicates the bosonic
representation the ket belongs to. It is to be noticed that at the right
hand side of Eq.(\ref{eq:basis}), only those states with the total number 
$\mathcal{N}$ fixed by the initial state are present. Using the
Hamiltonian in the diagonal form, Eq.(\ref{eq:Hnum}), and
Eq.(\ref{eq:basis}) the time-evolved state can be written as
\begin{eqnarray}
\ket{\psi(t)}&=&e^{-\imath \Phi} \sum^{n_1}_{i=0}\sum^{n_2}_{j=0}
c_{i,j}(n_1,n_2)e^{2 \imath (i+j)\lambda t} \times \nonumber \\
&&\ket{\mathcal{N}-(i+j),(i+j)}_A, \label{eq:psit1}
\end{eqnarray}
where the quantity $\Phi=\omega_0 t(\mathcal{N}+1)+\lambda
t\mathcal{N}+\hbar g t(\mathcal{N}+1)^2$ is a global phase. This phase
factor will be relevant in Section~\ref{sec:solcoe}, where this
result will be used to construct the solution for the case of
coherent states. Hence, for the number states, Eq.(\ref{eq:psit1}) shows
that the dynamics of the stationary states are completely
determined by the RWA coupling and the effect of non-linear part
of the Hamiltonian is only in the global phase.\\

After a little algebra we can re-write  the above state 
Eq.(\ref{eq:psit1}) in a more instructive form:
\begin{eqnarray}
\begin{array}{c}
\ket{\psi(t)}=\hat{\Gamma}^{(n_1)}_{12}(t) \,\hat{\Gamma}^{(n_2)}_{21}(t)\,
\ket{0,0}, \\ \\
\hat{\Gamma}^{(n)}_{kk'}(t) \equiv \frac{\left(\hat{a}^{\dag}_k
\cos{\lambda t}-\imath \hat{a}^{\dag}_{k'} \sin{\lambda t}
\right)^n}{\sqrt{n!}}.
\end{array}
\label{eq:numb}
\end{eqnarray}
The sub-indexes ``12''(``21'') of the operators $\hat{\Gamma}$
emphasize the entanglement features of the dynamics. Here, we omitted 
the remaining global phase, and used
the fact that the vacuum state in both representations spaces must
be the same.\\
\indent Now, we can extract some informations of
entanglement process of number states under the action solely of the RWA
interaction. At times $T_l=\frac{\pi}{\lambda}(l+\frac{1}{2})$,
with $l$ integer, Eq.(\ref{eq:numb}) gives
\begin{eqnarray}
\ket{\psi(T_l)}&=&(-\imath)^{n_1+n_2}
(-1)^{l(n_1+n_2)}\ket{n_2,n_1}, \label{eq:enumtpi2}
\end{eqnarray}
indicating that the system is in a disentangled state, but not
the same as the initial one. However, at times $\tau_l=l \frac{\pi}{\lambda}$ 
we recover the exact initial state (modulo an overall phase):
\begin{eqnarray}
\ket{\psi(\tau_l)}=(-1)^{l(n_1+n_2)}\ket{n_1,n_2}.
\label{eq:enumtpi}
\end{eqnarray}
Hence, for the case of initially disentangled number
states, we have found two characteristic times: (a)
the {\it re-coherence times} $T_l$, for the which the subsystems
recover the purity of the initial state, and (b) the {\it
recurrence times} $\tau_l$, when the evolved state
becomes equal to the initial state. This allows us to
classify $\ket{\psi(0)}$ in Eq.(\ref{eq:ininumber}) as a {\it reversible state} 
for this particular interaction. Consequently, we may define a 
{\it period of reversibility} for
such states: $\tau_{R}=\frac{\pi}{\lambda}$. For the special case
of {\em equal} initial number states ($n_1=n_2$),  the difference
between the re-coherence and recurrence times disappears and the
reversibility occurs earlier, as can be seen in
Eq.(\ref{eq:enumtpi2}).
%
\subsection{\label{sec:solcoe} Analytical solution for coherent states}
Consider an initially disentangled product of two coherent states as follows:
\begin{eqnarray}
|\psi(0)\rangle&=&|\alpha_1\rangle\otimes|\alpha_2\rangle\\
&=&e^{\frac{-|\alpha_1|^2}{2}}e^{\frac{-|\alpha_2|^2}{2}}\sum_{n,m}
\frac{\alpha_1^n}{\sqrt{n!}}
\frac{\alpha_2^m}{\sqrt{m!}}|n,m\rangle. \nonumber \label{eq:icohst}
\end{eqnarray}
The time-evolved state can be obtained using the previous result
Eq.(\ref{eq:numb}), including the $\mathcal{N}$-dependent phase $\Phi$
\begin{eqnarray}
|\psi(t)\rangle
&=&e^{\frac{-|\alpha_1|^2-|\alpha_2|^2}{2}}\sum_{n,m}
\frac{\alpha_1^n(t)}{\sqrt{n!}} \frac{\alpha_2^m(t)}{\sqrt{m!}} e^{-\imath 
\hbar g \left(n+m +1\right)^2 t} \times \nonumber \\
&& \times\,\Gamma^{(n)}_{12}(t)
\,\Gamma^{(m)}_{21}(t)\,|0,0\rangle, \label{eq:cohst}
\end{eqnarray}
where $\alpha_k(t)=\alpha_ke^{-\imath(\omega_0+2\hbar g)t}$. In this 
result we have already omitted a global phase.

It is easy to see that for $g = 0$ the RWA-coupling does not
entangle the oscillators, since the two summations of
Eq.(\ref{eq:cohst}) can be factorized, and each oscillator remains
in a coherent form
\begin{eqnarray}
|\psi(t)\rangle |_{g=0}&=&\hat{D}_1[\beta_1(t)]\hat{D}_2
[\beta_2(t)]|0,0\rangle\nonumber\\
&=&|\beta_1(t),\beta_2(t)\rangle, \label{eq:rwa}
\end{eqnarray}
where
\begin{eqnarray}
\beta_{1 \atop 2}(t)=(\alpha_{1 \atop 2}e^{-\imath\omega_0
t}\cos{\lambda t}- \imath\alpha_{2 \atop 1}e^{-\imath\omega_0
t}\sin{\lambda t}), \label{eq:beta}
\end{eqnarray}
and $\hat{D}_k[\beta_k(t)]=e^{\left(\beta_k \hat{a}^\dagger_k -
\beta_k^* \hat{a}_k\right)}$ is the displacement operator in the
phase space of the $k$-th oscillator.

Using this particular result, we can solve the general case of
initially disentangled coherent states in a more 
intuitive way. The commutation relations (\ref{eq:commutator})
allow us to apply separately the piece of the evolution operator 
associated to the non-linear term of $\hat{H}$, Eq.~\ref{eq:hamiltonian},
and use the previously derived results [Eqs.(~\ref{eq:cohst},~\ref{eq:rwa})]. The final exact solution 
for the temporal evolution of two initially coherent states is
given by
\begin{eqnarray}
\left|{\psi(t)}\right\rangle&=& e^{ \frac{g t \hat{H}_0^2 }
{\imath \hbar} } e^{\frac{\omega t \hat{H}_0+\lambda t
\hat{V}}{\imath\hbar}}|\alpha_1,
\alpha_2\rangle\nonumber \\
&=& e^{ \frac{g t \hat{H}_0^2 }{\imath \hbar} }|\beta_1(t),\beta_2(t)
\rangle\nonumber\\
&=&e^{ - \frac{{\left| {\beta_1 \left( t \right)} \right|^2  +
\left| {\beta_2 \left( t \right)} \right|^2 }}{2}}
\sum\limits_{n,m}
e^{ - \imath \hbar g t{\cal N}^2} \times \nonumber \\
&\times &{\frac{{\left[ {\beta_1 (t) e^{-\imath 2 g \hbar t}}
\right]^{n} }}{{\sqrt {n!} }}}\frac{{\left[ {\beta_2 (t)e^{-\imath
2 g \hbar t}} \right]^{m} }}{{\sqrt {m!} }} \left| {n,m}
\right\rangle. \nonumber\\ \label{eq:psit2d}
\end{eqnarray}
Here, the expressions of $\beta_k(t)$ are the ones depicted in Eq.
(\ref{eq:beta}).

 In order to calculate other quantities like the mean values and variances 
of the quadrature operators, and discuss the phenomenon of 
collapses and revivals, let us calculate the density operator of the system. 
Since we are interested in the case where the global system is isolated, 
the total density operator is a projector onto the state $|{\psi(t)}\rangle$
\begin{eqnarray}
\hat\rho\left(t\right)& =& \left|\psi\left( t \right)
\right\rangle\left\langle\psi \left( t \right)\right| \nonumber\\
&=&e^{-\frac{\Lambda}{2\hbar}} \sum\limits_{n,{m}}
\sum\limits_{n',{m}'}
e^{-\imath\hbar g t\left[\left(n+{m}+1\right)^2-\left(n'+{m}'+1\right)^2
\right]}\nonumber\\
&\times & \frac{\beta_1^n \beta_1^{*n'} \beta_2^m
\beta_2^{*m'}}{\sqrt{n!n'!m!m'!}} |n,m\rangle \langle n',m'|,
\label{eq:DO}
\end{eqnarray}
where we have defined $\Lambda=\left|\beta_1 \right|^2+\left|
\beta_2\right|^2$, a constant of motion associated to the mean value
of $\mathcal{\hat{N}}$. We shall denote from here on $\beta_k(t)$ 
simply as  $\beta_k$. We calculated the reduced density operators 
$\hat{\rho}_k(t)$ ($k=1,2$) by tracing over the undesired degree 
of freedom, corresponding to one of the original oscillators:
\begin{eqnarray}
\hat \rho _{1}\left( t \right) &=&
e^{-\frac{\Lambda}{2\hbar}} \sum\limits_{n,m} \frac{{\beta_{1}^n }}
{{\sqrt {n!} }} \frac{{\beta_{1}^{*m}
}}{{\sqrt {m!} }} e^{|\beta_{2}|^{2}e^{-2\imath\hbar 
g t\left(n-m\right)}}\nonumber\\
&\times & e^{ - \imath \hbar g t\left[ {n^2  - m^2 + 2\left( {n -
m} \right)} \right]} \left| n \right\rangle \left\langle {m}
\right|, 
\label{eq:reducedDO}
\end{eqnarray}
with a similar expression for $\hat \rho _{2}\left( t \right)$.
The field quadrature operators $\hat{Q_k}$ and $\hat{P_k}$ are given in terms
of the creation and annihilation operators as follows: 
\begin{eqnarray} 
{\hat{Q}_k\choose \hat{P}_k}=\sqrt{\frac{\hbar}{2}} \left[ 
\begin{array}{cc} 
1 & 1 \\ -\imath & \imath 
\end{array} 
\right] {\hat{a}_k \choose \hat{a}^{\dag}_k}, 
\end{eqnarray} 
and analytical expressions for the mean values are given by
\begin{eqnarray} 
\left<\hat{Q}_k\right>(t)&=&Tr_k[\hat{Q}_k\,\hat{\rho}_k(t)] \\ 
&=&\sqrt{2\hbar} \,\, \textrm{Re}\left[\beta_k(t)e^{-3\imath \omega_g t} 
e^{-\frac{\Lambda}{2 \hbar} \left(1-e^{-2\imath 
\omega_g t}\right)}\right],\nonumber 
\label{eq:mvqi} 
\end{eqnarray} 
\begin{eqnarray} 
\left<\hat{P}_k\right>(t) &=& Tr_k[\hat{P}_k\,\hat{\rho}_k(t)]\\ 
&=& \sqrt{2\hbar}\,\, \textrm{Im}\left[\beta_k(t) e^{-3\imath \omega_g t} 
e^{-\frac{\Lambda}{2 \hbar} \left(1-e^{-2\imath \omega_g t} 
\right)}\right], \nonumber
\label{eq:mvpi} 
\end{eqnarray} 
where we defined the frequency associated to the non-linear term 
as $\omega_g=\hbar g$.
\begin{figure}[ht] 
\centerline{\includegraphics[scale=0.35]{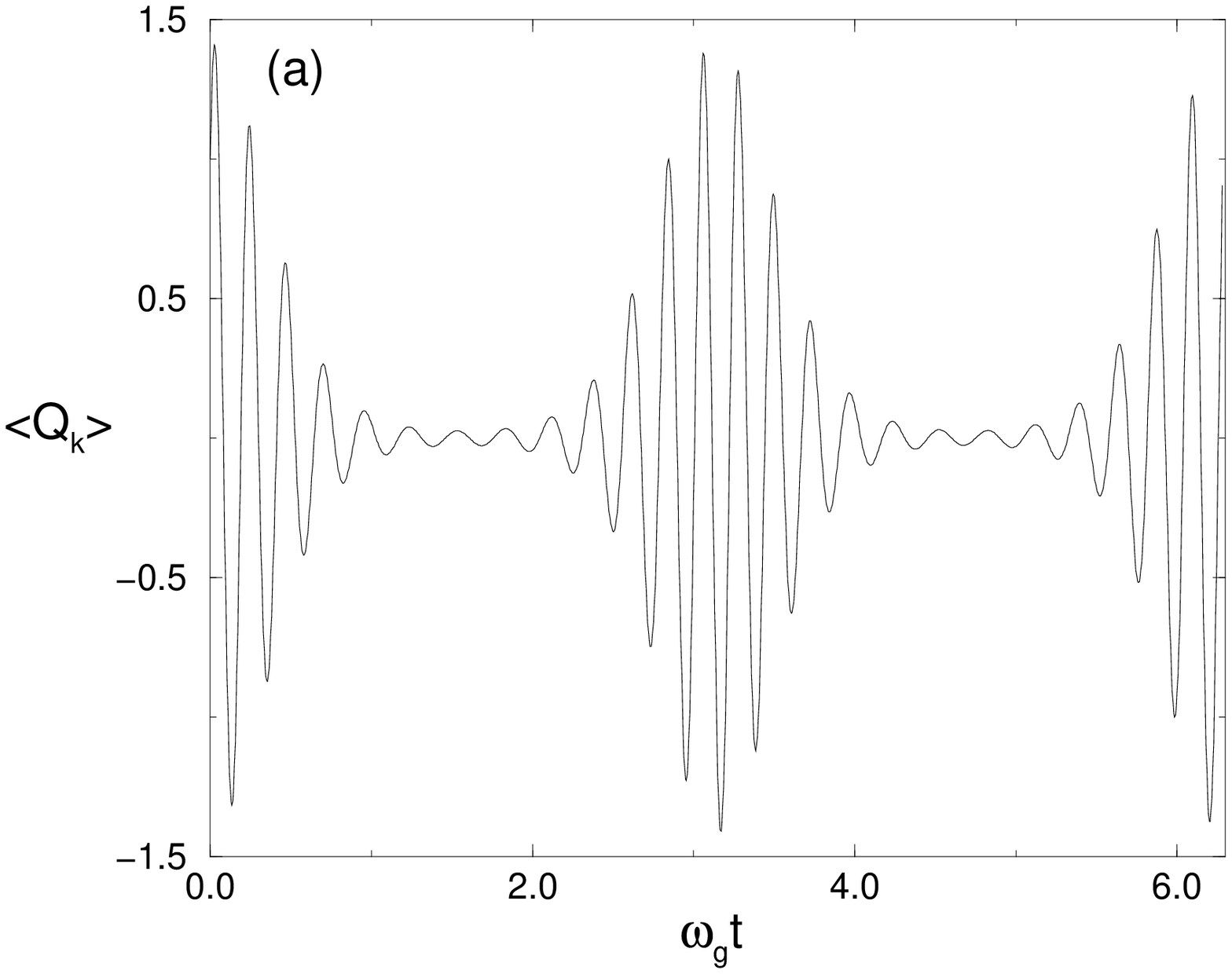}} 
\centerline{\includegraphics[scale=0.35]{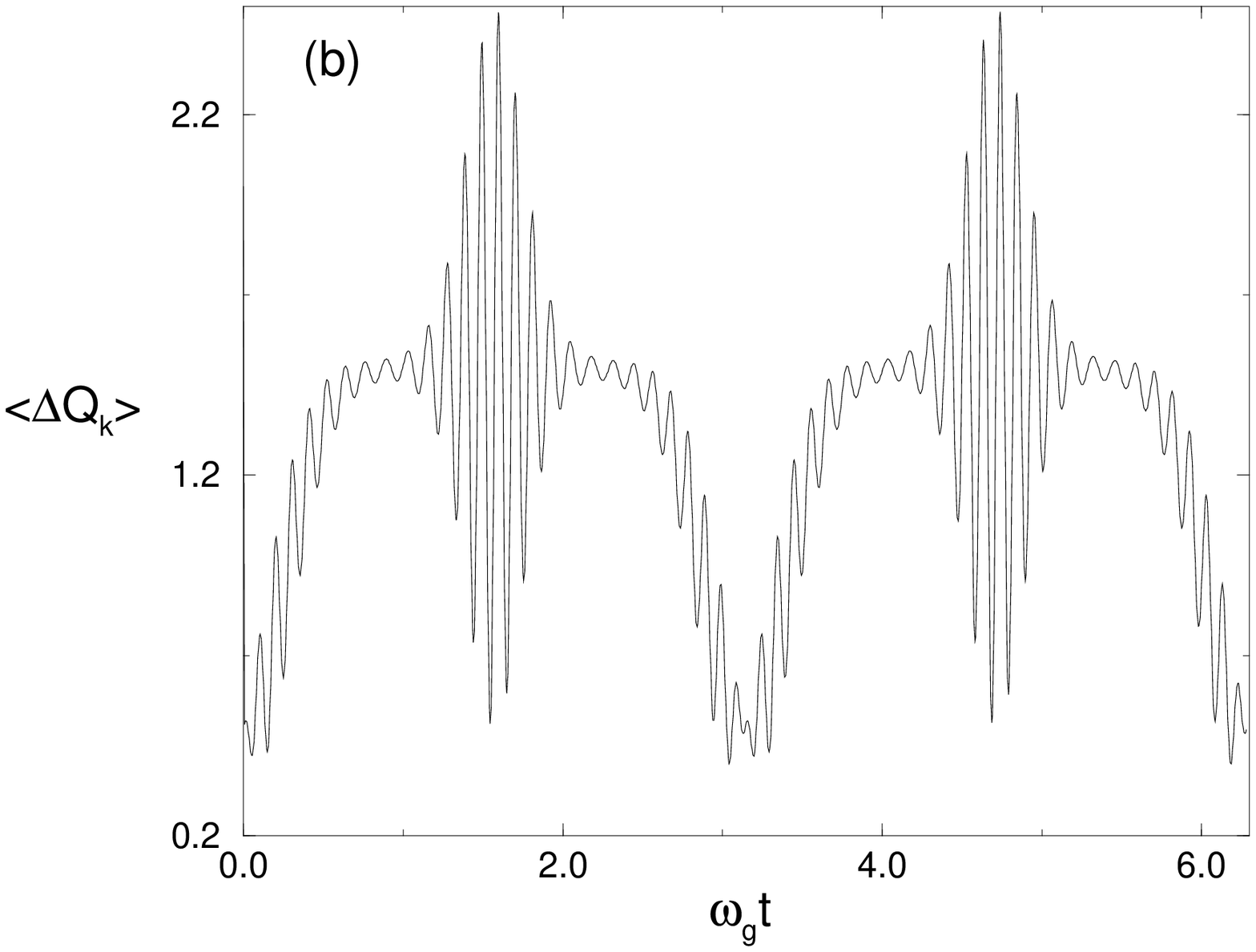}} \vspace{-0.5cm} 
\caption{\small 
(a) Collapse and revival phenomenon in the 
evolution of the expectation value of the quadrature $\hat{Q}_k$ 
and (b) the respective variance. Analytical results for 
$\frac{\omega_0}{\omega_g}= 20$, $\frac{\lambda}{\omega_g}=2$, 
 $q_{k0}=p_{k0}=1.0$, 
$\Lambda=4.0$ and $\hbar=1$. The same qualitative behaviors are 
observed for $\langle \hat{P}_k\rangle(t)$ 
and $\langle \Delta\hat{P}_k\rangle(t)$.
} 
\label{fig:revival} 
\end{figure} 
The above expressions contain exponentials of oscillatory terms, 
which produces the {\em collapses} and {\em revivals} 
phenomena \cite{agarwal89} at times odd multiple of $\frac{\pi}{2\omega_g}$
and integer multiples of $\frac{\pi}{\omega_g}$ respectively, as shown in 
Fig.\ref{fig:revival}(a) where we plotted the mean value of $\hat{Q}_k$ 
quadrature as a function of $\omega_g t$. 

One can also calculate analytically the {\em variances} 
$\langle\Delta Q_k\rangle(t)$ and  $\langle\Delta P_k\rangle(t)$. The 
results are complicated expressions which we shall omit. However, 
we illustrate its general behavior in Fig.\ref{fig:revival}(b), where 
we can see the same periodic structure present in the quadrature mean value 
temporal behavior. These results for the subsystem variances are very 
similar to those obtained for the one degree of freedom quartic oscillator 
\cite{milburn86}. The most instructive information is the 
relation between the collapse and the phase space's {\it delocalization in phase} 
of the state, as can be seen in the projected {\it Q}-function 
that will be illustrated in the next section for a particular set 
of parameters.

  Due to the fact that here we have two 
interacting subsystems, their entanglement prevent the appearance of 
superpositions of coherent states for each oscillator for low 
integer fractions of the re-coherence times. But, in contrast to the
case coupled to a reservoir \cite{milburn2}, we will see in what follows 
that at the revival time the subsystem entropy goes to zero (re-coherence), 
and the system do come back to the initial state (recurrence) under 
appropriate conditions. Another initially disentangled state that can be 
solved is the product of the form $\ket{\alpha_1}\otimes \ket{n_2}$, where 
this type of analysis can be done and find the times at re-coherence
allows superposition states  \cite{angelo}.
\section{\label{sec:entanglement}Entanglement properties and its 
semi-classical behavior.} 
In this section we are going to discuss the entanglement dynamics of 
time-evolved states found in Section~\ref{sec:model}; namely, the product
of number states and coherent states. We analyze the subsystem entropy, one 
important tool that gives a measure of entanglement for globally pure 
bipartite systems, and also the Husimi distribution 
or {\it Q}-function, in order to follow the evolution of the partial
distribution in phase space. The {\it Q}-function for the global system 
is usually defined by:
\begin{equation}
Q\left(\vec{q},\vec{p}\right)=\frac{1}{\pi}\bra{\gamma_1\,\gamma_2}
\hat{\rho}(t) \ket{\gamma_1\,\gamma_2},
\label{eq:Qfunction}
\end{equation}
where we choose $\gamma_k= \frac{q_k+\imath p_k}{\sqrt{2 \hbar}}$ ($k=1,2$).
Particularly, we investigate the role played by the non-linear 
interaction term in the Hamiltonian (\ref{eq:hamiltonian}) which appears
associated to the frequency $\omega_{g}$ in the various 
analytical expressions derived. Moreover, we would like to find the
behavior of the subsystem linear entropies in the {\it semi-classical limit}.  
\subsection{\label{sec:entnumber} Entanglement properties for number state: 
linear and Von Neumann entropies.} 
\begin{figure}[ht] 
\centerline{\includegraphics[scale=0.35]{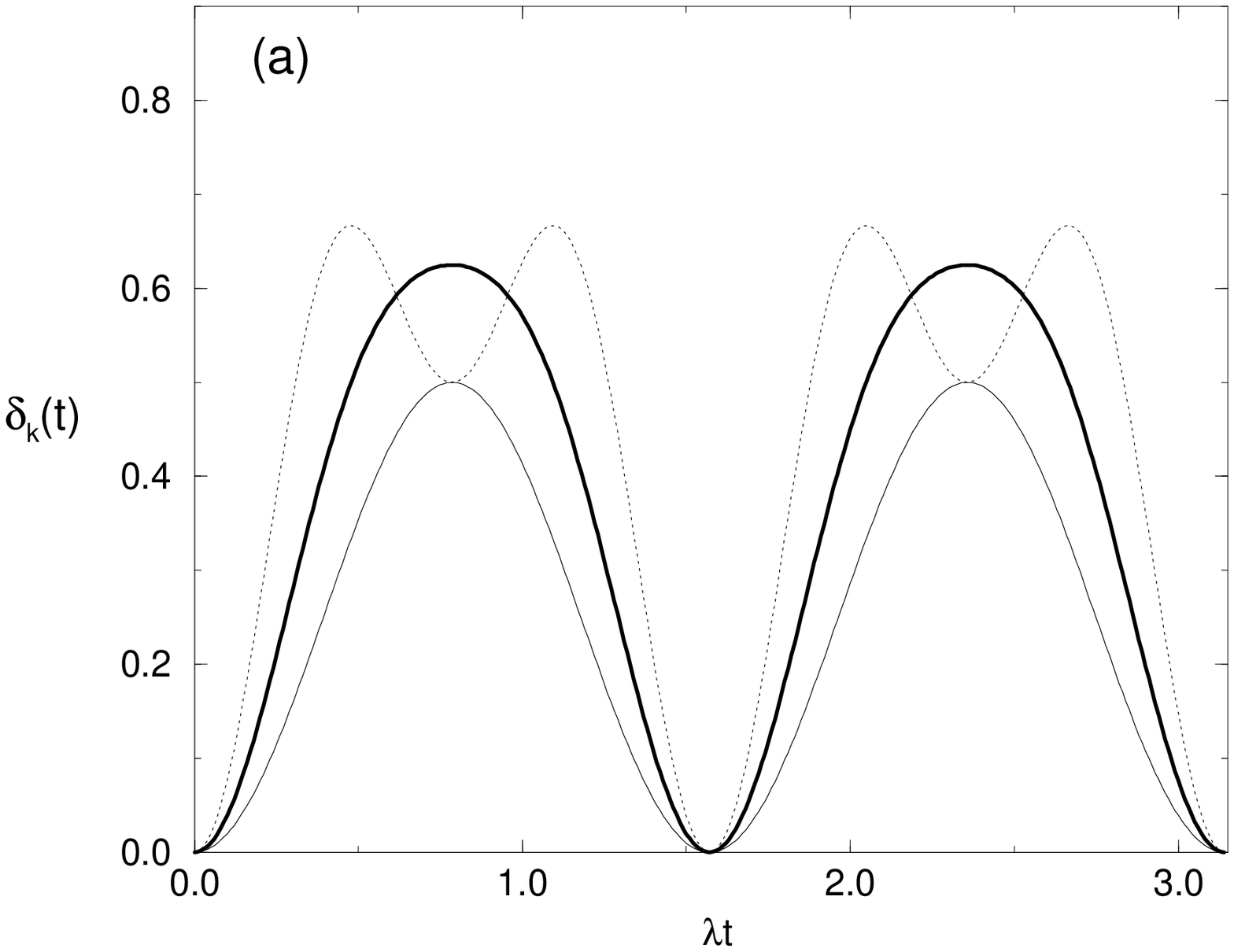}}
\label{fig:dn}
\centerline{\includegraphics[scale=0.35,angle=0]{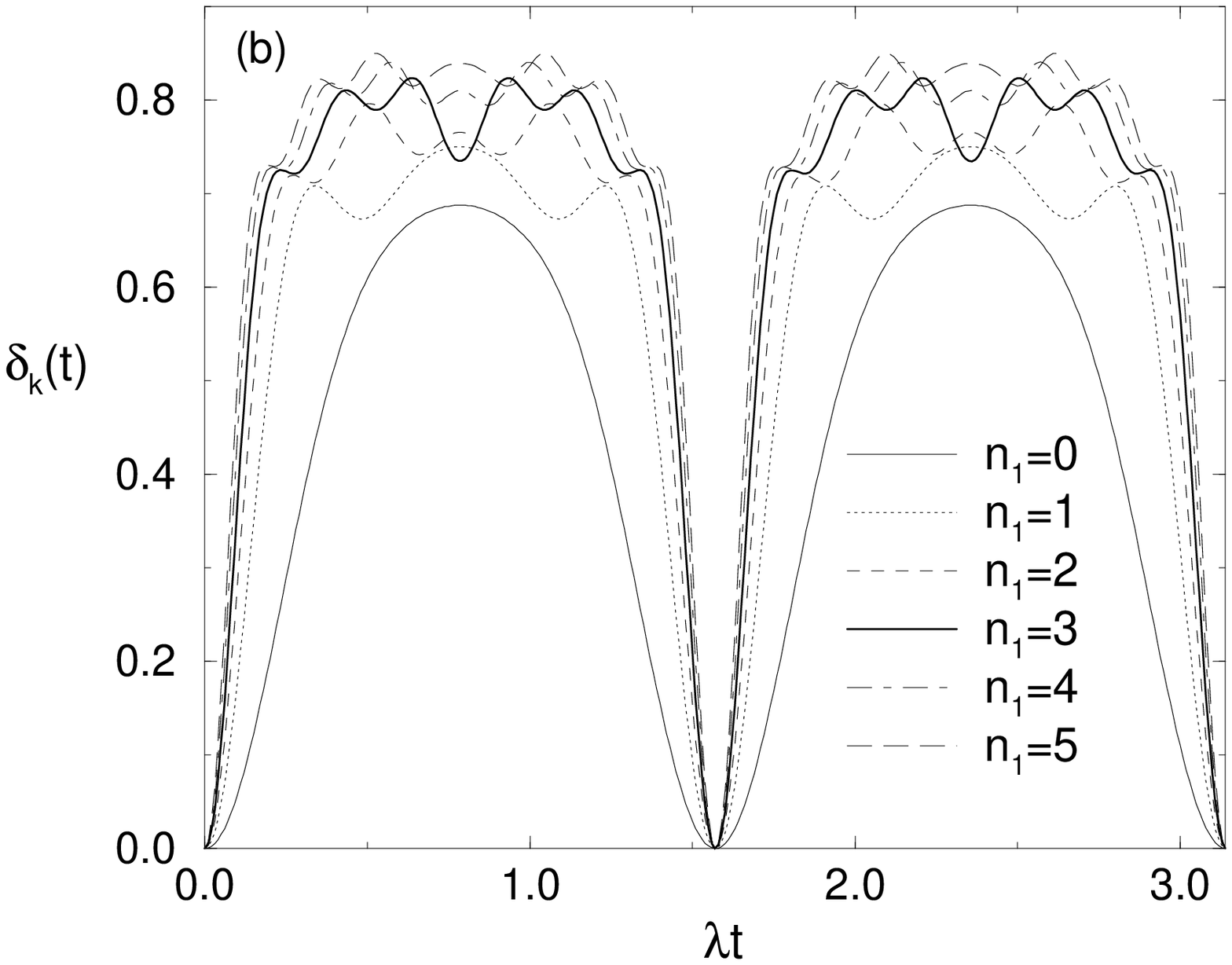}}
\vspace{-0.5cm}
 \caption{\small 
(a) Subsystem linear entropy for three different sets of initial number 
states: $\ket{1,0}$ (solid line), $\ket{1,1}$ (dashed line) and $\ket{2,0}$ 
(solid thick line).
(b) Subsystem linear entropy for initial number states ($|n_1,n_2\rangle$),
with $n_2=3$ and several values of $n_1$ (legends). Note the particular 
behavior for $n_1=n_2$ at $\lambda t=\frac{\pi}{2}$.
}
\label{fig:dn3}
\end{figure}

We are interested in the calculation of both, the subsystem linear entropy 
(SLE) and the Von Neumann entropy, usually defined by
\begin{eqnarray}
\delta_k&=&1-Tr_k(\hat{\rho}_k^2)\nonumber\\
S_k&=&-Tr_k(\hat{\rho}_k\ln{\hat{\rho}_k}),
\label{eq:defent}
\end{eqnarray}
where the label $k$ is associated to one of the degrees of freedom.   
For the initially disentangled  number states Eq.(\ref{eq:ininumber}), 
the reduced density operator is diagonal in the number state basis; 
therefore, it can be directly 
expressed in terms of its eigenvalues as follows 
\begin{equation} 
\rho_k(t)=\sum_{l=0}^{n_1+n_2}\lambda_l(t)|l\rangle\langle l|, 
\label{eq:rhoi} 
\end{equation} 
where the trace condition $\sum\lambda_l(t)=1$ must be 
satisfied. The eigenvalues $\lambda_l(t)$ are given by: 
\begin{equation} 
\lambda_l(t)=\sum_{i,i'=0}^{n_1}\sum_{m,m'=0}^{n_2} 
c_{i,m}(t)c^*_{i',m'}(t)\, 
\delta_{{i-m},{i'-m'}}\,\delta_{{n_1-l},{i-m}} \label{eq:eigenv} 
\end{equation}  
\begin{eqnarray} 
c_{i,m}(t)&=&{n_1 \choose i}{n_2 \choose 
m}\sqrt{\frac{(n_1-i+m)!(n_2-i+m)!} {n_1!n_2!}} \times 
\nonumber \\ 
&&\times(\cos{\lambda t})^{n_1+n_2-(i+m)}(-\imath \sin{\lambda 
t})^{i+m}. \label{eq:c} 
\end{eqnarray} 
Then, in terms of the reduced density operator eigenvalues, the 
subsystem entropies assume a familiar form: 
\begin{equation} \begin{array}{c} 
\displaystyle{\delta_k(t)=1-\sum_{l=0}^{n_1+n_2}\lambda_l(t)^2} \\ \\ 
\displaystyle{S_k(t)=-\sum_{l=0}^{n_1+n_2}\lambda_l(t) 
\ln{\left(\lambda_l(t) \right)}}. 
\end{array} 
\label{eq:entropies} 
\end{equation} 
The dependence of the eigenvalues on the periodic functions confirms the 
rule for the re-coherence times, Eq.(\ref{eq:enumtpi2}) and 
Eq.(\ref{eq:enumtpi}). To illustrate what has been discussed above, 
we present some simple examples: 
\begin{itemize} 
\item For $\ket{\psi(0)}=\ket{1,0}$ we have: 
{\small 
\begin{eqnarray}  \left\{ \begin{array}{l} 
\delta_k(t)=\frac{1}{2} \sin^2{2\lambda t}, \\ \\ 
S_k(t)=-\cos^2{\lambda t} \ln\left({\cos^2{\lambda t}}\right)- 
\sin^2{\lambda t} \ln\left({\sin^2{\lambda t}}\right). 
\end{array} \right. 
\label{eq:ds10} 
\end{eqnarray}} 
\item For $\ket{\psi(0)}=\ket{1,1}$ we have: 
{\small 
\begin{eqnarray} \left\{ \begin{array}{l} 
\delta_k(t)=\frac{\sin^2{2 \lambda t}}{4} \left(5+3\cos{4\lambda t} 
\right), \\ \\ 
S_k(t)=-\cos^2{2\lambda t} \ln\left({\cos^2{2\lambda t}}\right)- 
\sin^2{2\lambda t} \ln\left({\frac{\sin^2{2\lambda t}}{2}}\right). 
\end{array} \right. 
\label{eq:ds11} 
\end{eqnarray}} 
\item For $\ket{\psi(0)}=\ket{2,0}$ we have: 
{\small 
\begin{eqnarray} \left\{ \begin{array}{l} 
\delta_k(t)=\frac{\sin^2{2 \lambda t}}{16} \left(13+3\cos{4\lambda 
t} \right), \\ \\ 
S_k(t)=-\cos^4{\lambda t} \ln\left({\cos^4{\lambda t}}\right)- 
\sin^4{\lambda t} \ln\left({\sin^4{\lambda t}}\right)+ \\ \\ 
-\frac{\sin^2{2\lambda t}}{2} \ln\left({\frac{\sin^2{2\lambda 
t}}{2}}\right). 
\end{array} \right. 
\label{eq:ds20} 
\end{eqnarray}} 
\end{itemize} 
These analytical results for the SLE are shown in Fig.\ref{fig:dn3}(a). 
An interesting fact is the appearance of the oscillatory structures 
between two successive re-coherences for particular values of $n_k$.
This is illustrated in Fig.\ref{fig:dn3}(b) where the SLE is plotted for 
 $n_1=3$ and several successive values of  $n_2$. We can see that both,
the number of oscillations and the maximum value of SLE, increases as  
$n_2$ is increased. This feature can be understood from Eq.(\ref{eq:rhoi}), 
where the number of accessible states is given by $n_1+n_2+1$ which increases 
with increasing $n_2$. This accounts for the increase in the maximum
value, and also the subsystem is allowed to pass through many more possible 
mixed states.  
Notice that no dependence on $\hbar$ or $g$ is essentially left in the 
expression of the time-evolved state (\ref{eq:psit1}), and the same happens 
to the calculated SLE's [Eqs.(\ref{eq:ds10}-\ref{eq:ds20})]. This is the 
case where the non-linear interaction plays no
role, and the semi-classical limit is related to the presence of many 
photons (large $n_k$). 
%
\subsection{\label{sec:entcoe}Entanglement properties for coherent states 
and semi-classical behavior.}
\begin{figure}[ht]
\centerline{\includegraphics[scale=0.4]{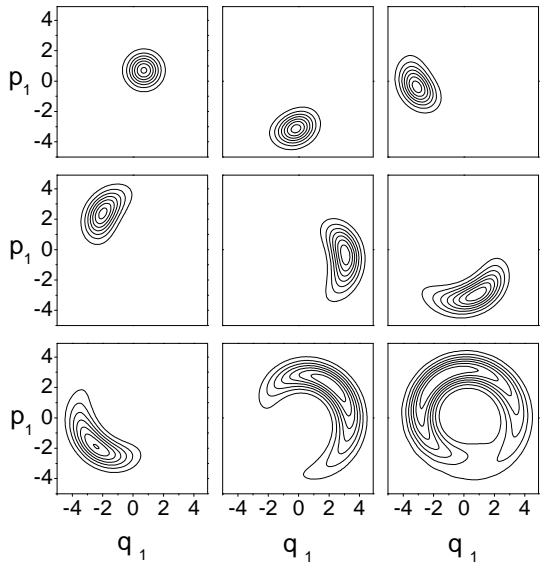}}
\vspace{-0.5cm}\caption{\small Contour plot of subsystem {\it Q}-function 
on plane ($q_1,p_1$) during the {\it phase spread regimen}, for initial conditions 
$q_{10}= p_{10}= q_{20}= p_{20}= 1.0$, $\omega_0=1$, $\lambda=0.2$, $g =0.1$ 
and $\mathcal{R}= 0.025$. The first plot correspond to the initial time 
(upper-left corner) and the subsequent times are chosen in the interval 
$0 < t \le t_b$.
}
\label{fig:husimi1}
\end{figure}

\begin{figure}[ht]
\centerline{\includegraphics[scale=0.4]{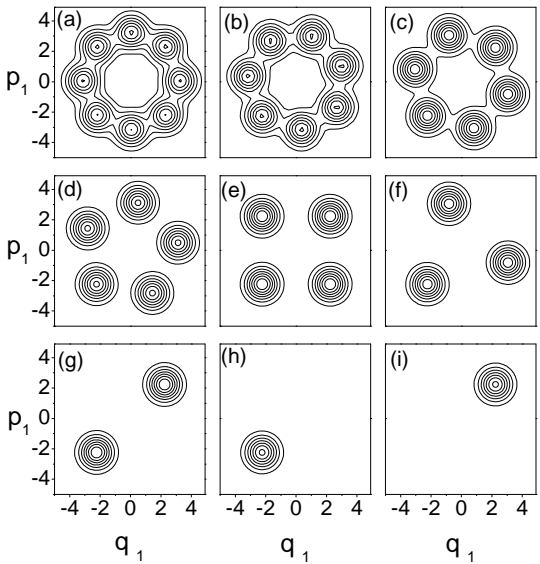}}
\vspace{-0.5cm}\caption{\small Time evolution of contour plots of a subsystem
{\it Q}-function during the {\it self-interference regimen} for $\mathcal{R}= 0.025$
in the plane ($q_1,p_1$), with the same parameter values of 
Fig.~\ref{fig:husimi1}. (a) $t=T_1/8$,
(b)$t=T_1/7$, (c) $t=T_1/6$, (d) $t=T_1/5$, (e) $t=T1R/4$, (f)$t=T_1/3$, 
(g) $t=T_1/2$, (h) $t= T_1$ (re-coherence) , (i) $t=2 T_1=\tau_1$ 
(recurrence). 
}
\label{fig:husimi2}
\end{figure}

In order to study the entanglement dynamics and reversibility properties 
for the initial product of coherent states, we calculate the SLE defined 
in Eq.(~\ref{eq:defent}). Using previous result Eq.(\ref{eq:psit2d}), 
an exact expression for this quantity can be calculated:
\begin{eqnarray}
\delta_{1}(t)&=&1-Tr_{1}\big(\hat{\rho}^2_{1}(t)\big) 
\nonumber \\
&=& 1 -e^{ -2|\beta_{1}\left(t\right)|^2}
\sum\limits_{n,m}
\frac{|\beta_{1}\left(t\right)|^{2n}}{n!}\frac{|\beta_{1}
\left(t\right)|^{2m}}{m!} \nonumber \\
&\times& e^{-4|\beta_{2}\left(t\right)|^2\sin^2{\left[\hbar g t
\left(n-m\right)\right]}}.
\label{eq:rle}
\end{eqnarray}
In a similar fashion as in the case of the number states, we can find
the re-coherence times studying the conditions under which $\delta_{1}(t)$
is equal to zero, indicating that the subsystems are disentangled. A simple
analysis of Eq.(\ref{eq:rle}) shows that this happens in the following 
situations:
\begin{enumerate}
\item For any initial conditions excluding the vacuum state ($\ket{0,0}$), 
the re-coherence times associated to the non-linear interaction are
$\hbar gT_l=l\pi$, when the argument of the sine-function is an integer 
multiple of $\pi$:
\begin{equation} 
T_l=l\frac{\pi}{\hbar g},\, \mbox{($l>0$ integer)}.
\label{trecoh}
\end{equation}
\item For those initial conditions such that one of the arguments 
$\beta_k(t)=0$, one can find other instants when SLE is zero. These 
conditions are associated exclusively with RWA coupling and 
special initial conditions. In terms of the quadratures of the initial 
values $\alpha_k$, using the expression (\ref{eq:beta}) one can
write these conditions as follows:\\
\centerline{$q_{1}q_{2}+p_{1}p_{2}=0$,}
and we have disentangled states for times \cite{19}:\\

\centerline{$ t_1=\frac{1}{\lambda}arctan(-\frac{q_1}{p_2})$,
or $t_2=\frac{1}{\lambda}arctan(\frac{p_1}{q_2})$.}
\end{enumerate}
Among these re-coherence times, we can identify those at which we also 
have recurrences by examining Eq.(\ref{eq:psit2d}). Only the first class of
re-coherences produces recurrences, and this happens when the following 
condition is satisfied:
{\em for a given value of $T_l$, when the ratio  $l\frac{\omega_0}{\omega_g}$ and 
$l\frac{\lambda}{\omega_g}$ are integer numbers.}\\
%
\begin{figure}[ht]
\centerline{\includegraphics[scale=0.45]{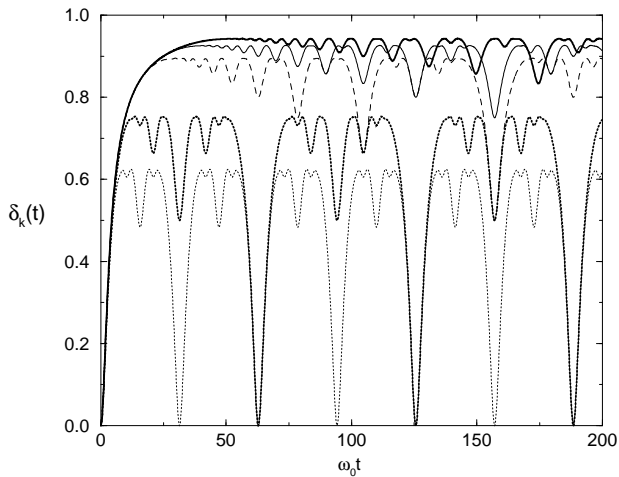}}
\vspace{-0.5cm} 
\caption{\small
Subsystem linear entropy  $\delta_k$ as function of $\omega_0t$
for $\frac{\lambda}{\omega_0}=0.2$,
$\frac{\omega_g}{\omega_0}=0.1$, $q_{k0}=p_{k0}=1.0$, $\Lambda=4$
and several values of $\mathcal{R}=\frac{\hbar}{\Lambda}$: $\mathcal{R}=
\frac{1}{4}$ (dotted line), $\mathcal{R}=\frac{1}{8}$ (dotted thick line),
 $\mathcal{R}=\frac{1}{40}$ (dashed line), $\mathcal{R}=\frac{1}{80} 
$(solid line) and $\mathcal{R}=\frac{3}{400}$  (solid thick line).
}
\label{fig:entropy}
\end{figure}
Now we will choose a particular parameter sets to illustrate the semi-classical
limit of the SLE, since in contrast to the case of the number states, we 
have an explicit dependence on $\hbar$, actually on the frequency 
$\omega_g= \hbar g$. This allows us to study the semi-classical behavior of the
SLE. The natural parameter to measure the `quantumness' of the system here 
is the ratio ${\cal R}=\frac{\hbar}{\Lambda}$, 
where $\Lambda$ (defined in Section \ref{sec:solcoe}) is a characteristic 
action in phase space. In Fig.\ref{fig:entropy} we plotted the time evolution 
of the SLE for several values of ${\cal R}$. Here, we adopted the convention
to fix the value of $\Lambda$ and vary $\hbar$ instead of the opposite way.
The first thing to be noticed
is the fact that all curves coincides at the short time scale, where the
SLE increases monotonically until it reaches the maximum value. The maximum
value of SLE depends on ${\cal R}$, which increases as we let ${\cal R} \ll 
1$. This has to do with the increasing number of accessible states as 
we let the spectrum become denser, and also implies in loss of information.
We will call this first regimen the `{\em phase spread regimen}' in connection to 
the behavior of the {\it Q}-function of each subsystem that we shall see in 
what follows. After reaching the maximum value, oscillations start to happen 
in the SLE, which can be seen as a partial recovering of coherence, until 
the first re-coherence time $T_1=\frac{\pi}{g\hbar}$ (see Eq.(\ref{trecoh})) 
at which $\delta_k=0$ and the subsystem recovers purity. This second regimen 
will be called  `{\em self-interference regimen}' since the time-evolved subsystem
{\it Q}-function shows the phenomenon of self-interference typical of the
Kerr-type nonlinearity. This is consistent with the
fact that in the limit ${\cal R} \rightarrow 0$ the re-coherence
time goes to infinity and the initial purity will never be recovered.
We will call {\em break time} $t_b$, the time at which the transition from the 
phase spread regimen (rising) to the self-interference (oscillating) regimen 
occurs. It is clear from Fig.\ref{fig:entropy} that this time increases with 
the inverse power of ${\cal R}$, and we also expect it to go to infinity in 
the classical limit \cite{20}.\\
Let us illustrate the differences between the two regimens by means of 
the subsystem {\it Q}-function which will show the proper signatures in each 
regimen. In Fig.~\ref{fig:husimi1} we plotted {\it Q}-function in various 
instants of the {\em phase spread regimen} for the same Hamiltonian parameters 
used in Fig.~\ref{fig:entropy}, and the quantumness parameter ${\cal R}=0.025$
. The sequence of plots is in the quadrature plane of the  oscillator-$1$, 
beginning at $t=0$ until $t \approx t_b$. Where the center of the initially
coherent Gaussian wave-packet follows essentially the classical trajectory, 
as predicted by the  Ehrenfest theorem \cite{ehrenfest}, circulating around
the origin and, due to the nature of the self-interaction term, the packet 
itself spreads in phase angle in the phase space. During the interval of time 
before the front of the packet reaches its tail \cite{milburn2},
no re-coherences can happen.\\
In Fig.\ref{fig:husimi2}, another sequence of contour plots of 
{\it Q}-functions of the subsystem-$1$ shows a time evolution during the
{\it self-interference regimen} in the interval of time after the break time $t_b$,
until the first re-coherence time $T_1$. We also add a plot at the recurrence 
time ($\tau_1=2T_1$ for this particular set of parameters). It is remarkable 
the appearance of several peaks along the annular 
region, a signature of self-interference phenomenon where a kind of standing 
waves with $M$-peaks forms at times $t=T_1/M$ (for $t>t_b$),  when the SLE
assumes the value $\delta_k\approx 1-\frac{1}{M}$ at a local minimum.  
This kind of behavior is a hallmark of the self-interference regimen, an 
essentially quantum phenomena which is at the core of re-coherence and 
reversibility in this system.
This sequence of phenomena {\em phase spread} and {\it self-interference} is 
reproduced many times latter since the entanglement process in this case is 
reversible. We call the attention to the fact that in this particular case 
shown, the standing waves are {\em not} Schr\"odinger cat states, in fact it
is a mixture of cat states, in contrast to the case reported in Ref.
\cite{agarwal98,banerji} for a similar Kerr-type Hamiltonian. We present this 
calculation in the Appendix \ref{app:mixture}.
\section{\label{sec:conclusions} Conclusions}
We have solved analytically the problem of two resonant RWA-interacting fields
in the presence of non-linear Kerr-like interactions.
The time-evolved state was exactly determined and 
this allowed us to identify analytically how the collapses and revivals are 
produced in the quadrature mean values for the initially coherent state.
All properties of the entanglement dynamics have been studied for initial 
states which are products of both coherent and number states.

In particular, we presented all the necessary conditions to the re-coherence 
of the initially non-entangled number and coherent states.  We have calculated
the exact expressions for the subsystem entropies in both cases.
 Also, some conditions for the recurrence has been established for the coherent
state case,  related to the commensurability of the physical frequencies of 
the model. 

We also identified, in the case of coherent initial states, two distinct 
regimens of entanglement: the first one ({\it phase spread regimen}) happens
during the time where the initial coherent state spreads in phase angle
in the phase space; whereas the second one ({\it self-interference regimen}) occurs
when the phase spread state starts to self-interfere. The time at which
self-interference becomes important for the evolution of this type of initial
states, we call it {\it break time} ($t_b$), and it delimits the   
beginning of the essentially quantum processes responsible for the 
re-coherence. We also have shown that the self-interference of each oscillator
produces the standing waves, where no single Schr\"odinger cat state
is allowed but a mixture of Schr\"odinger cat states, consistent with the
entanglement.
Finally, we show how in the semi-classical limit ($\hbar \rightarrow 0$)
both the re-coherence times and the recurrence times go away and the 
entanglement process becomes irreversible for all practical purposes in 
this type of model for Gaussian initial states.

\begin{acknowledgments}
The authors acknowledge K. F. Romero for helpful discussions 
and Conselho Nacional de Desenvolvimento Cient\'{\i}fico e
Tecnol\'ogico (CNPq) (Contracts No.300651/85-6,No.146010/99-0) and 
Funda\c{c}\~ao de Amparo \`a Pesquisa de S\~ao Paulo (FAPESP)
(Contract No.98/13617-4) for financial support.
\end{acknowledgments}
\appendix
\section{\label{app:mixture} A demonstration of mixture of Schr\"odinger 
cat states}
We first re-write Eq. (\ref{eq:psit2d}) in a more compact notation
\begin{eqnarray}
\left|{\psi(t)}\right\rangle&=&\sum\limits_{n,m}
c_n\left(\gamma_1\right) c_m\left( \gamma_2\right)
e^{ - \imath g \hbar t (n+m)^2} \left| {n,m} \right\rangle
\label{psizip}
\end{eqnarray}
\begin{eqnarray}
c_n(\gamma)=e^{-\frac{|\gamma|^2}{2}}\frac{\gamma^n}{\sqrt{n!}}
\end{eqnarray}
where $\gamma_k=\beta_k e^{-\imath 2 g \hbar t}$ and $\beta_k$ are 
functions of time. We are interested in the instants 
$t_{r,s}=\frac{\pi}{g \hbar}\frac{r}{s}=T_1 \frac{r}{s}$, where $r$ and 
$s$ are mutually prime with $r<s$. Using the discrete Fourier 
transform 
\begin{eqnarray}
e^{-\imath \pi n^2 \frac{r}{s}}=\sum^{l-1}_{q=0}a_q^{(r,s)}e^{-\imath 2 \pi 
n \frac{p}{l}}
\end{eqnarray}
where
\begin{eqnarray} l=\left\{
\begin{array}{l}
s,\,\,\textrm{if $r$ is odd and $s$ is even, or vice-versa,} \\
2s,\,\,\textrm{if both $r$ and $s$ are odd;}
\end{array} \right.
\end{eqnarray}
and
\begin{eqnarray}
a_q^{(r,s)}=\frac{1}{l}\sum^{l-1}_{k=0}e^{-\imath\pi k \left(k\frac{r}{s}-2
\frac{q}{l} \right)},
\end{eqnarray}
we re-write Eq.(\ref{psizip}) at the mentioned 
instants as follows:
\begin{eqnarray}
\left|{\psi(t_{r,s})}\right\rangle&=&\sum\limits_{n,m}
c_n\left(\gamma_1\right) c_m\left( \gamma_2\right)
e^{ - \imath \pi \frac{r}{s} (n+m)^2} \left| {n,m} \right\rangle \nonumber \\ 
&=& \sum^{l-1}_{q,p=0}a_q a_p \sum\limits_{n,m} c_n\left(\eta_{1q} 
\,e^{-\imath2\pi m \frac{r}{s}}\right) c_m\left(\eta_{2p}\right)
\left| {n,m} \right\rangle \nonumber \\
\end{eqnarray}
\begin{eqnarray}
\qquad\nonumber\\
&=& \sum^{l-1}_{q,p=0}a_q a_p \sum\limits_{m}
\left| \eta_{1q} \,e^{-\imath2\pi m \frac{r}{s}} \right\rangle 
\otimes  c_m\left(\eta_{2p} \right) |m\rangle
\nonumber \\
&=& \sum_m \left|\mathcal{C}_m \right\rangle \otimes
\sum^{l-1}_{p=0}a_p c_m\left(\eta_{2p} \right) |m\rangle\,.
\end{eqnarray}
Here $\eta_{kq}=\gamma_k e^{-\imath 2 \pi \frac{q}{l}}= 
\beta_k e^{-\imath 2 \pi \left( \frac{q}{l}+\frac{r}{s} \right)}$, and
we have defined the following cat-like states for the oscillator-$1$\\
\begin{eqnarray}
\left|\mathcal{C}_m\right\rangle&=&
\sum^{l-1}_{q=0}a_q \left| \eta_{1q} \,e^{-\imath2\pi m \frac{r}{s}} 
\right\rangle .
\end{eqnarray}
 Then, constructing the global density operator and tracing 
over the oscillator-$2$, i.e., summing over the left over number states, 
we finally get the following mixture of cat states:
\begin{eqnarray}
\rho_1(t_{r,s})&=&\sum_m \xi_m
\left| \mathcal{C}_m \right\rangle \left\langle \mathcal{C}_m \right| 
\label{mix} \\ \nonumber \\
\xi_m&=&\sum_{p,p'=0}^{l-1} a_p a^{*}_{p'} c_m(\eta_{2p}) c^{*}_m(\eta_{2p'}).
\end{eqnarray}


\begin{thebibliography}{20}
\bibitem{18} This choice of parameters is important when one seeks a 
well defined classical counterpart for the quantum Hamiltonian whose 
dynamics will be associated in the limit $\hbar \rightarrow 0$.
\bibitem{19} A more detailed study about this case where Shr\"odinger
cat states can be generated will be given in \cite{angelo}.
\bibitem{20} The connection of the break time with the Ehrenfest time
will be given elsewhere.
\end{thebibliography}
\end{document}